\documentclass[aps,prb,twocolumn]{revtex4-1}

\usepackage{graphicx}
\usepackage{color}
\usepackage{graphicx}
\usepackage{amssymb}
\usepackage{epstopdf}
\usepackage{times}
\usepackage{ctable}
\usepackage{booktabs}
\begin{document}


\title{Effective screening and the plasmaron bands in Graphene} 



\author{Andrew L. Walter$^{1,2,a}$, Aaron Bostwick$^{1}$, Ki-Joon Jeon$^{3}$, Florian Speck$^{4}$, Markus Ostler$^{4}$, Thomas Seyller$^{4}$, Luca Moreschini$^{1}$, Young Jun Chang$^{1,2}$, Marco Polini$^{5}$, Reza Asgari$^{6}$, Allan H. MacDonald$^{7}$, Karsten Horn$^{2}$, Eli Rotenberg}

\address{ $^1$Advanced Light Source (ALS), E. O. Lawrence Berkeley National Laboratory, Berkeley, California 94720, USA,\\ $^{2}$ Department of Molecular Physics, Fritz-Haber-Institut der Max-Planck-Gesellschaft, Faradayweg 4-6, 14195 Berlin, Germany,\\ $^{3}$School of Electrical Engineering, University of Ulsan, Namgu, Ulsan, 680-749, South Korea\\ $^{4}$ Lehrstuhl f\"{u}r Technische Physik, Universit\"{a}t Erlangen-N\"{u}rnberg, Erwin-Rommel-Strasse 1, 91058 Erlangen, Germany.\\$^{5}$NEST, Istituto Nanoscienze-CNR and Scuola Normale Superiore, I-56126 Pisa, Italy.\\$^{6}$School of Physics, Institute for Research in Fundamental Sciences (IPM), Tehran 19395-5531, Iran.\\ $^{7}$Dept. of Physics, Univ. of Texas at Austin, 1 University Station C1600, Austin TX 78712 \\ $^{a}$ email: alwalter@lbl.gov}



\date{\today}

\def\EF{$E_{\mathrm{F}}$}
\def\ED{$E_{\mathrm{D}}$}
\def\Ea{$E_{\mathrm{0}}$}
\def\Eb{$E_{\mathrm{1}}$}
\def\Ec{$E_{\mathrm{2}}$}
\def\kF{$k_{\mathrm{F}}$}
\def\ky{$k_{\mathrm{y}}$}
\def\kx{$k_{\mathrm{x}}$}
\def\kp{$k_{\parallel}$}
\def\kk{$K-K$}
\def\gk{$K-\Gamma$}
\def\AG{$\alpha_{\mathrm{G}}$}
\def\dk{$\delta k$}
\def\dE{$\delta E$}
\def\EBind{$E_{\mathrm{B}}$}
\def\RPA{$G_{0}W$ -RPA}

\begin{abstract}
Electron-plasmon coupling in graphene has recently been shown to give rise to a "plasmaron" quasiparticle excitation. The strength of this coupling has been predicted to depend on the effective screening, which in turn is expected to depend on the dielectric environment of the graphene sheet. Here we compare the strength of enviromental screening for graphene on four different substrates by evaluating the separation of the plasmaron bands from the hole bands using Angle Resolved PhotoEmission Spectroscopy. Comparison with $G_{0}W$-RPA predictions are used to determine the effective dielectric constant of the underlying substrate layer. We also show that plasmaron and electronic properties of graphene can be independently manipulated, an important aspect of a possible use in "plasmaronic" devices.

\end{abstract}


\maketitle


\section{Introduction}

Bohm and Pines\cite{Bohm:1953p4724} first discussed the role of interactions between charged quasiparticles and the collective charge density oscillations of the electron gas (plasmons) in the renormalization of the electronic band structure. Lundqvist\cite{Lundqvist:1967p4797} later predicted that  the electrons, holes and plasmons could couple to form a new quasi-particle known as a plasmaron. These plasmarons have been observed experimentally by optical\cite{Shay:1971p4812,Tediosi:2007p4817}, tunnelling\cite{DialIII:2007p4853} and angle resolved photoemission\cite{Bostwick:2010p3387} spectroscopies. In studying the valence electronic structure of graphene, side bands due to plasmaron excitation are accessible to experimental investigation in great detail \cite{Polini:2008p3589}. The dependence of these features on electron density confirms quantitatively, by comparison to $G_{0}W$-RPA theory, that these are in fact due to electron-plasmon coupling. The resulting charge density waves are expected to oscillate at much higher frequencies than hole charge density waves\cite{Rana:2008p4661,RYZHII:2006p4660} providing a motivation for the development of "plasmaronic" devices, which merge photonics and electronics. A fundamental requirement of these devices is the ability to manipulate the electronic and plasmaronic properties independently. Graphene, on account of its simple and well understood band structure, provides the ideal system to investigate these issues\cite{Polini:2008p3589,Hwang:2008p3441,Bostwick:2010p3387}. 

The dependence of the plasmaron coupling strength on the effective screening of graphene by the underlying substrate has been investigated theoretically\cite{Polini:2008p3589}, however no experimental work has been published. Here we present Angle Resolved Photoemission Spectroscopy (ARPES) measurements of graphene on SiC with four different interface layers: $(6\sqrt{3}\times6\sqrt{3}) R30^{\circ}$ C-SiC (henceforth $6\sqrt{3}$ C-SiC), Au-SiC, F-SiC and H-SiC. These are expected, from previous experimental results, to exhibit considerable differences in the effective screening of the graphene. We characterize the effect of differing substrates on the plasmaronic bands and show that in all cases the spectral function of n-type graphene depends only on a single, substrate dependent, coupling constant \AG\ regardless of the doping level. Analysis of the measurements indicate a range of substrate effective dielectric constants ($\epsilon_{\mathrm{S}} \sim 87$ to $\sim 7.8$) indicating that, in addition to many other unique properties, graphene is ideally suited to investigate the role of substrate screening on the band structure.

In the first hexagonal carbon layer formed by thermal decomposition of the SiC(0001)\cite{Seyller:2006p4029,Emtsev:2008p4975,Emtsev:2009p212} surface, every third atom is covalently bonded to the substrate. This creates a $(6\sqrt{3}\times6\sqrt{3}) R30^{\circ}$ C layer that does not show the linear band structure characteristic of graphene. The subsequent layers formed by further thermal decomposition are decoupled from the substrate/ underlying $6\sqrt{3}$ C layer and exhibit the characteristic bandstructure of graphene\cite{Ohta:2007p258}. Previously hydrogen\cite{Riedl:2009p4922,Speck:2010p5373,Bostwick:2010p3387,Virojanadara:2010p5374} , fluorine\cite{Walter:2011gf} and gold\cite{Gierz:2008p3404} have been shown to intercalate under layers of $6\sqrt{3}$ C on SiC, decoupling it from the substrate and transforming it into graphene. While the single graphene layer on $6\sqrt{3}$ C- SiC is n-type with the Dirac crossing, \ED\ , at $\sim$ - 0.45 eV \cite{Bostwick:2007p247} graphene formed by intercalation is p-type with \ED\ at 0.15 eV (Au-SiC), 0.18 eV (H-SiC) and 0.79 eV (F-SiC). To investigate the plasmaronic bandstructure, these intercalated systems were converted to n-type by chemical doping with K atoms.


\section{Background}
\label{sec.background}
\begin{figure}
\includegraphics{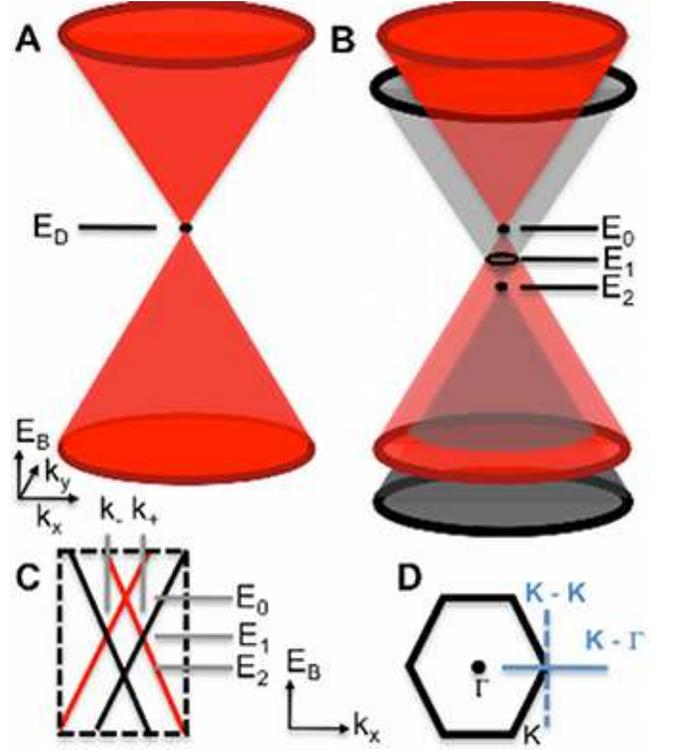}
\caption{\label{Dirac_cones} The linear band dispersion in; \bf A\rm, the non-interacting single particle (bare band) description, and \bf B\rm, the case of interactions between elementary charges and plasmons. In \bf C \rm a linear cut of the Dirac region in \bf B \rm (\ky\ = 0) is shown. hole bands are shown in red while plasmaron bands are shown in black. The Dirac energy crossing, \ED\ , in \bf A \rm is split into three crossings by the interactions with the plasmons; \Ea\ , the hole band crossing, \Ec\ , the plasmaron band crossing, and \Eb\ the ring like crossing of the hole and plasmaron bands.The scaled momentum and energy widths of the diamond region in \bf C \rm are given by $\delta k = | k_{\mathrm{+}}-k_{\mathrm{-}}|/|k_{\mathrm{F}}|$ and $\delta E = | E_{\mathrm{2}}-E_{\mathrm{0}}|/|E_{\mathrm{0}}-E_{\mathrm{F}}|$ respectively. In \bf D \rm a plot of the graphene Brillouin zone, showing the principle points and principle directions referenced in the paper, is presented.}
\end{figure}

The, n-type, Dirac cone picture of the electron dispersion near the Fermi energy, \EF\ , in graphene is shown in Fig. \ref{Dirac_cones} \bf A \rm for the non-interacting, single-particle description. In this case the bands are linear and form two cones which meet at a single point at the Dirac energy, \ED\ . This picture does not take into account the coupling between the elementary charges and the plasmons which leads to the creation of a second plasmaronic dispersion\cite{Polini:2008p3589,Hwang:wh,Hwang:to,Hwang:2008p3441,Bostwick:2010p3387}. The new picture is schematically shown in Fig. \ref{Dirac_cones} \bf B \rm where the hole dispersion, shown in red, is intersected by a second (plasmaronic) dispersion, shown in black. The Dirac crossing is split into three distinct crossings: the hole band crossing at \Ea\ , the plasmaron band crossing at \Ec\ and the ring like crossing of the hole and plasmaron bands at \Eb\ .The Dirac energy, \ED\ , is the crossing energy of the hole bands in the hypothetical non- interacting picture which can not be determined from the ARPES data directly. To overcome this in the following we will use the crossing of the electron- hole bands in the interacting case, \Ea\ , as an approximation. 

\begin{figure*}
\includegraphics{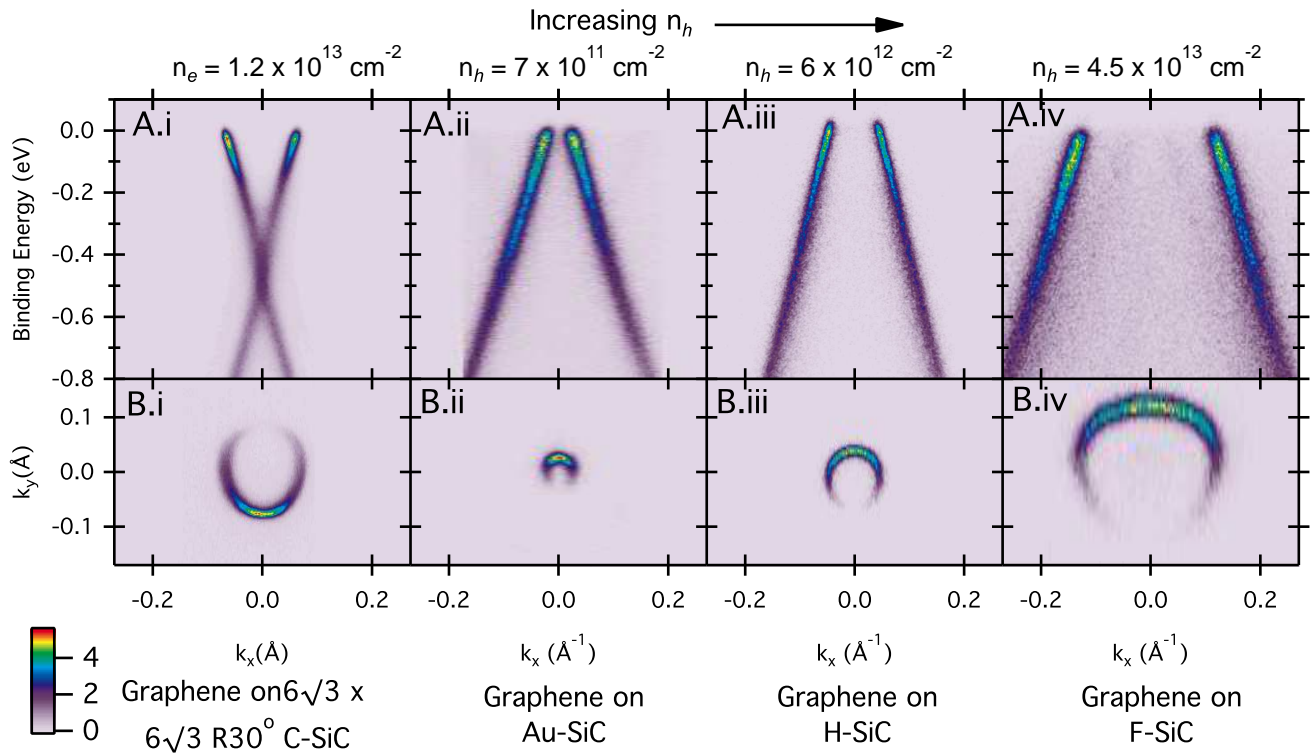}
\caption{\label{ARPES_before}Experimental spectral function of graphene on  $(6\sqrt{3}\times6\sqrt{3}) R30^{\circ}$ C-SiC, i, graphene on Au-SiC, ii, graphene on H-SiC, iii, and Graphene on F-SiC, iv, in the \kk\ direction, \bf A\rm, and the Fermi surface, \bf B\rm. The graphene on  $(6\sqrt{3}\times6\sqrt{3}) R30^{\circ}$ C-SiC sample is n-doped by the substrate (n$_e \approx 1.2$ x $10^{13}$ cm$^{-2}$) while the graphene on Au-SiC (n$_h \approx 7$ x $10^{11}$ cm$^{-2}$), graphene on H-SiC (n$_h \approx 6$ x $10^{12}$ cm$^{-2}$) and Graphene on F-SiC (n$_h \approx 4.5$ x $10^{13}$ cm$^{-2}$) are all p-doped by the substrate.}
\end{figure*}

\begin{figure*}
\includegraphics{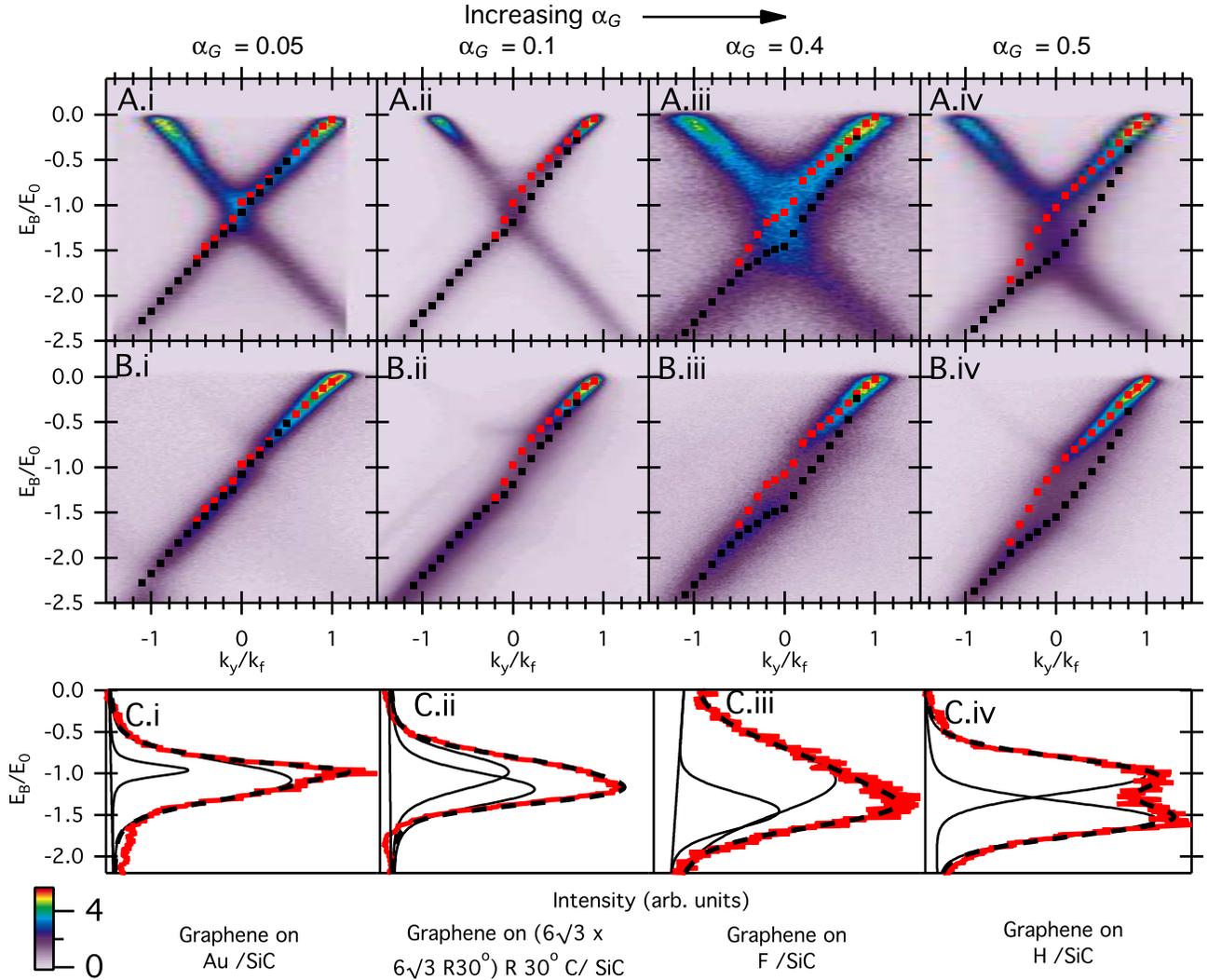}
\caption{\label{ARPES_after}Experimental spectral function of  graphene on Au-SiC, i, graphene on $(6\sqrt{3}\times6\sqrt{3}) R30^{\circ}$ C-SiC, ii, graphene on F-SiC, iii, and graphene on H-SiC, iv, in the \kk\ direction, \bf A\rm, and the \gk\ direction, \bf B\rm. As opposed to \ref{Dirac_cones} the order is now in terms of increaseing \AG\ . All samples are n-doped to n$_e \approx 6$ x $10^{13}$ cm$^{-2}$ by the addition of K atoms to the surface. All of the plots are presented with the energy scaled to the Dirac energy ($E_\mathrm{scaled} = E_\mathrm{B}/E_{\mathrm{0}}$) and the momentum scaled to the Fermi vector ($k_\mathrm{scaled} = k/|k_\mathrm{F}|$). The results of the energy line profile peak fitting to the electron hole bands and the plasmaron bands are shown in red and black respectively, while the $k/|k_\mathrm{F}| =0$ line profile and fitted peaks are shown in \bf C\rm . The separation of the hole and plasmaron bands increases from left to right and in all cases two bands (hole and plasmaron) are observed with the the hole(plasmaron) band showing greater intensity above (below) the Dirac crossings.}
\end{figure*}

\begin{figure}
\includegraphics{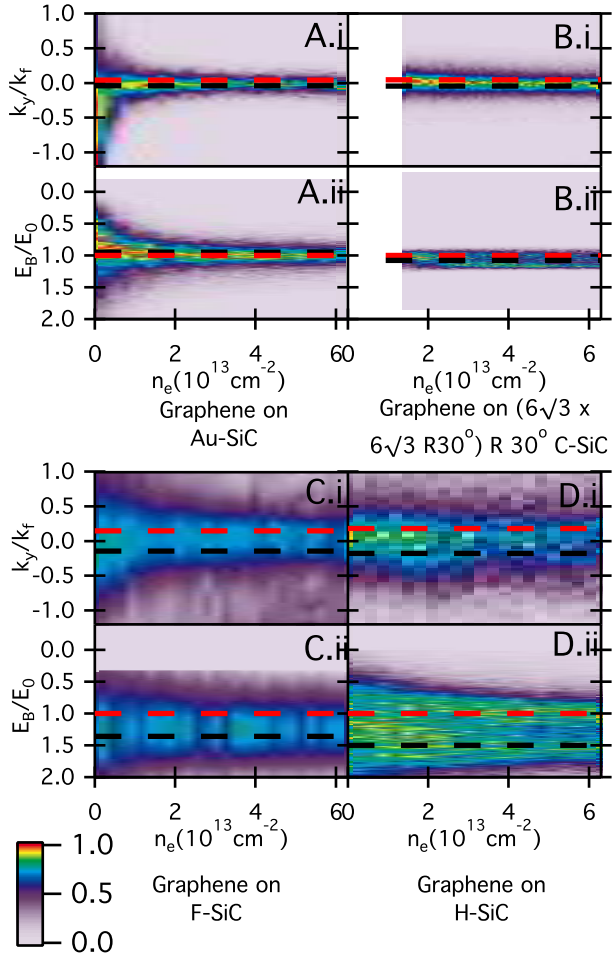}
\caption{\label{EDC}Constant energy ( \Eb\ ), i, and constant momentum ($k$ = 0), ii, intensity profiles as a function of K doping for  Graphene on Au-SiC, \bf A\rm, Graphene on $(6\sqrt{3}\times6\sqrt{3}) R30^{\circ}$ C-SiC, \bf B\rm, Graphene on F-SiC, \bf C\rm, and Graphene on H-SiC, \bf D\rm, in the \gk\ direction. All of the plots are presented with the energy scaled to the Dirac energy ($E_\mathrm{scaled} = E_\mathrm{B}/E_{\mathrm{0}}$) and the momentum scaled to the Fermi vector ($k_\mathrm{scaled} = k/|k_\mathrm{F}|$).The momentum ( \dk\ ) and Energy ( \dE\ ) separations in i and ii are constant except close to the zero doping level, where the vanishing density of states act to converge the hole and plasmaron bands.}
\end{figure}

\begin{figure}
\includegraphics{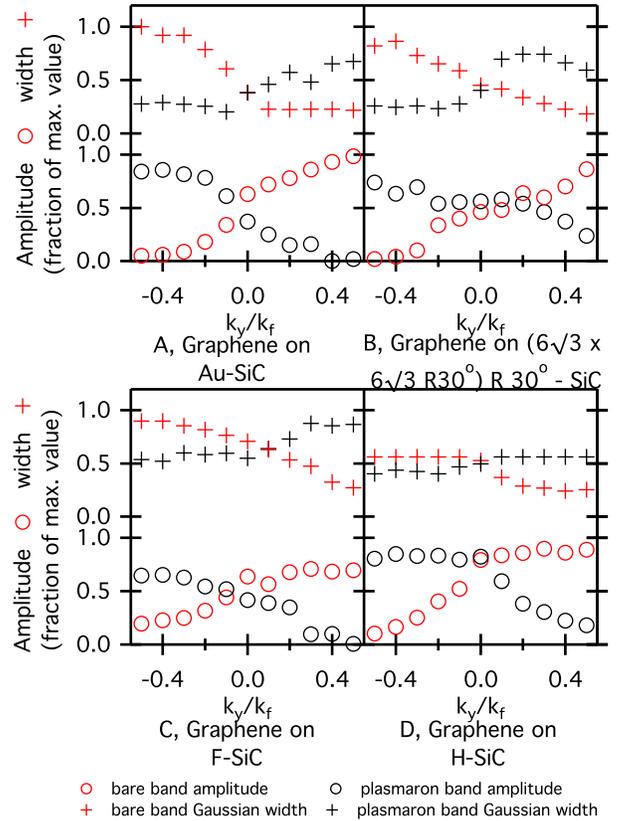}
\caption{\label{Fit_params}Peak amplitudes and Gaussian widths from the peaks shown in Fig. \ref{ARPES_after} for Graphene on Au-SiC, \bf A\rm, Graphene on $(6\sqrt{3}\times6\sqrt{3}) R30^{\circ}$ C-SiC, \bf B\rm, Graphene on F-SiC, \bf C\rm, and Graphene on H-SiC, \bf D\rm, in the \gk\ direction. In all cases we observe a similar increase (reduction) in the amplitude (Gaussian line width) of the hole band peaks, shown in red. The reverse is true for the plasmaron band peaks, shown in black. The fitted positions from Fig. \ref{ARPES_after} clearly show that for \ky\ / \kF\ $ < 0$ (\ky\ / \kF\ $ > 0$) the positions are below (above) the Dirac crossing.}
\end{figure}
 
\begin{figure}
\includegraphics{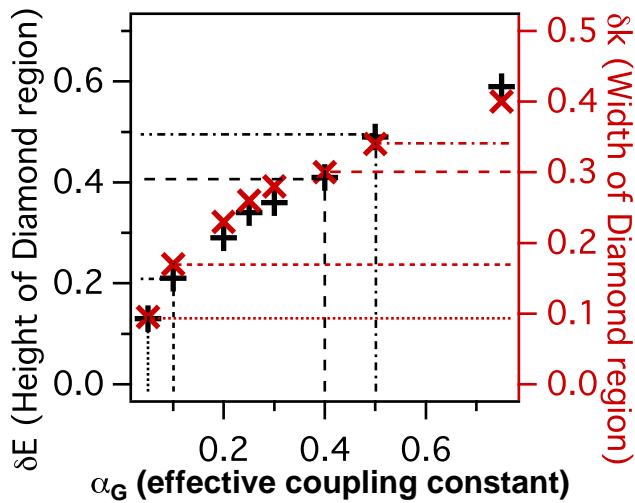}
\caption{\label{alpha}The energy separations, \dE\ , (in black) and momentum separations, \dk\ , (in red) of the "diamond" region as a function of graphene effective dielectric constant, \AG\ , as determined from one particle Green's function calculations within the Random Phase Approximation ( \RPA\ )\cite{Polini:2008p3589}. The values for \AG\ that correspond to the experimental widths shown in Table \ref{eff_screen} are indicated by dashed lines.}
\end{figure}

\begin{figure*}
\includegraphics{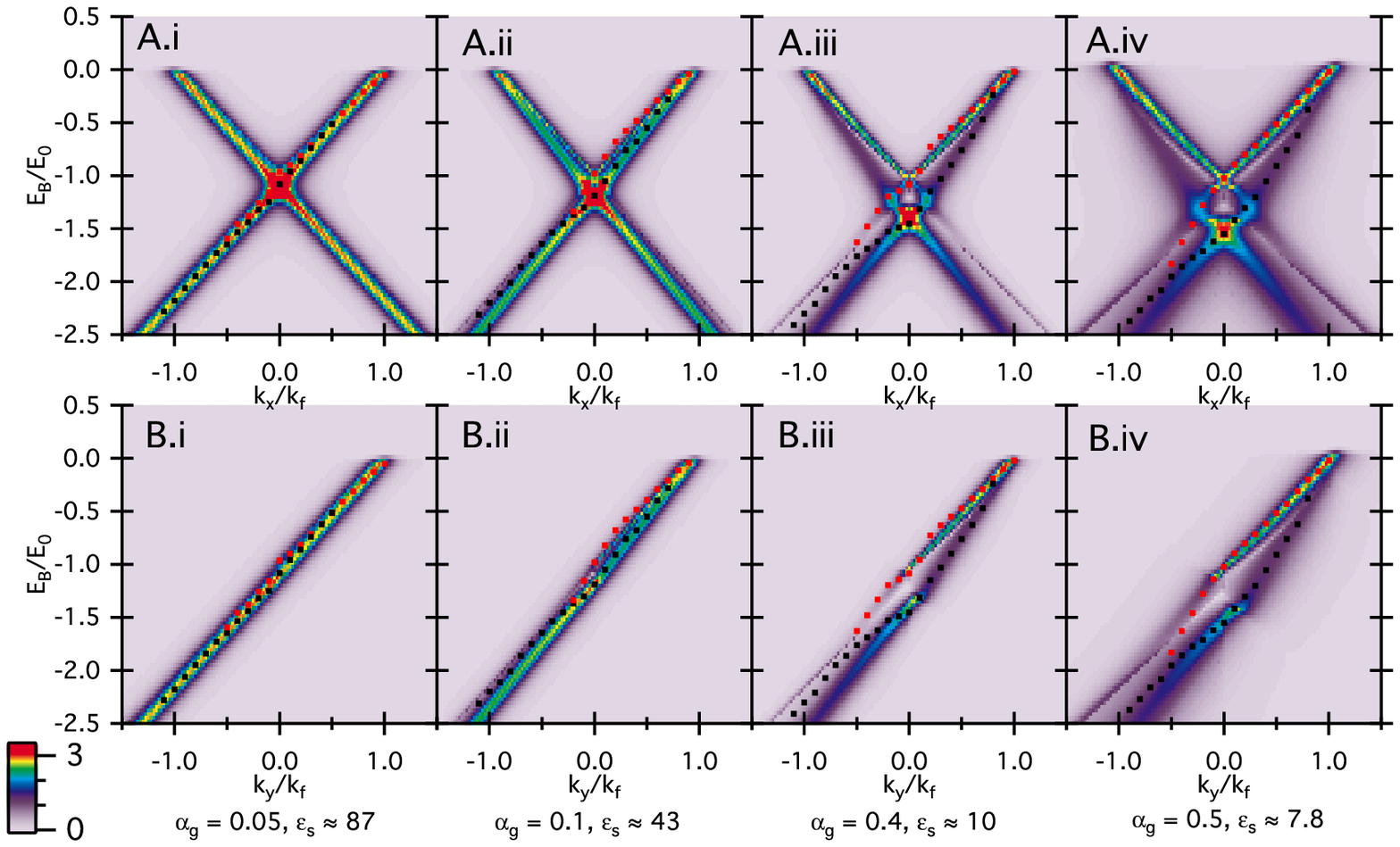}
\caption{\label{alpha_theory}One particle Green's function calculations within the Random Phase Approximation ( \RPA\ )\cite{Polini:2008p3589} of the spectral function of graphene in the \kk\ ( \bf A \rm) and \gk\ ( \bf B \rm) directions for the effective coupling constants ($\alpha _{\mathrm{G}}= e^2/ \epsilon \hbar v_{\mathrm{F}} \sim 2.2/\epsilon$ ) found experimentally. Overlayed are the experimental fitted positions from Fig.\ref{ARPES_after} for Graphene on Au-SiC, \bf i\rm, Graphene on $(6\sqrt{3}\times6\sqrt{3}) R30^{\circ}$ C-SiC, \bf ii\rm, Graphene on F-SiC, \bf iii\rm, and Graphene on H-SiC, \bf iv\rm. Above the lower Dirac crossing, \Ec\ , good agreement between the experiment and theory is observed, however below \Ec\ the hole (red) and plasmaron (black) bands quickly hybridize, deviating from the experimental prediction. The intensity at high binding energy is also over-estimated by the theory (see Fig. \ref{ARPES_after}).}
\end{figure*}

The linear band structure of graphene leads to a linear scaling of the Random Phase Approximation (RPA) dielectric function, and hence the momentum and energy separation of the hole and plasmaron bands, with \kF\ and \ED\ \cite{Polini:2008p3589,Hwang:wh,Hwang:to,Hwang:2008p3441,Bostwick:2010p3387}. This separation can be determined directly from ARPES data by a line shape analysis of the spectral function in terms of a hole band (top) and plasmaron (bottom) band. When taking an energy vs. momentum cut ($E(k_{x})$, at $k_{y}$ = 0) the bands form a "diamond" as seen schematically in Fig. \ref{Dirac_cones} \bf C \rm from which the scaled energy separation, \dE\ = $| E_{\mathrm{2}}-E_{\mathrm{0}}|/|E_{\mathrm{0}}-E_{\mathrm{F}}|$, and the scaled momentum separation, \dk\ = $| k_{\mathrm{+}}-k_{\mathrm{-}}|/|k_{\mathrm{F}}|$, can be determined.

In the absence of phonons, the plasmaronic band is predicted to merge with the hole band, and loses spectral weight close to the Fermi level, \EF\ . This is because the electrons only couple strongly to plasmons with the same velocity, a condition that cannot be met close to \EF\ .  The plasmon and the hole band travel with different velocity, greatly reducing the composite particle's lifetime.  With phonons, the situation is more complicated, since the bare plasmon breaks up into two modes \cite{Tegenkamp:2010vw,Hwang:2010th,Jablan:2009bs} . It is expected that this breakup will affect the spectral function near \EF\ but to our knowledge these effects have not been calculated to date.

In the one-particle Green function, Random Phase Approximation ( \RPA\ ) theory developed by Polini et al. \cite{Polini:2008p3589} and Hwang et al.\cite{,Hwang:wh,Hwang:to,Hwang:2008p3441} the effective environmental screening of the graphene layer, $\epsilon$, is related to a graphene effective coupling constant, $\alpha _{\mathrm{G}}$, via the relation $\epsilon = e^2/ \alpha _{\mathrm{G}} \hbar v_{\mathrm{F}} \sim 2.2/\alpha _{\mathrm{G}}$. For clarity note that $\epsilon$ is different from the RPA dynamical screening function, $\epsilon(q,\omega)$,  and that $v_{\mathrm{F}} \sim 1 \times 10^{6} $ m/s is the Fermi velocity. Increasing \AG\  decreases the RPA dielectric function and therefore increases the separation of the hole (upper) and plasmaron (lower) bands\cite{Polini:2008p3589,Bostwick:2010p3387}. The effective environmental screening of the graphene layer, $\epsilon$, is related to the dielectric constant from the substrate, $\epsilon_{\mathrm{S}}$, and the dielectric constant from the vacuum above, $\epsilon_{\mathrm{vac}} \approx 1$, via the relation $\epsilon = (\epsilon_{\mathrm{S}}+\epsilon_{\mathrm{vac}})/2 \Rightarrow \epsilon_{\mathrm{S}} \approx 2\epsilon - 1$. In reality this theory neglects remote band contributions to screening in the graphene layer and the polarizability of the K atoms used to electron dope the graphene, and therefore what is found is an upper limit to the dielectric constant.

\section{Experimental}
Photoemission spectra were obtained from graphene on $6\sqrt{3}$ C-SiC(0001) samples prepared by the high pressure Argon method, of Emtsev et al. \cite{,Emtsev:2009p212}and Ostler et al. \cite{Ostler:2010p5375}, and $6\sqrt{3}$ C-SiC(0001) samples\cite{Ostler:2010p5375} that were annealed in the presence of H, F or Au. Hydrogen\cite{Riedl:2009p4922,Speck:2010p5373,Bostwick:2010p3387,Virojanadara:2010p5374}, fluorine\cite{Walter:2011gf} and gold\cite{Gierz:2010cq} have been shown to intercalate under the $6\sqrt{3}$ C layer, decoupling it from the SiC substrate such that the $\pi $ electrons are free to form the linear bands characteristic of graphene. The hydrogen intercalation process follows the method of Riedl et al. \cite{Riedl:2009p4922} with $6\sqrt{3}$ C-SiC(0001) samples annealed at $\sim$570 $^{\circ}$C for $\sim$ 75 minutes in H$_{2}$. Treatment of the $6\sqrt{3}$ C-SiC(0001) with fluorine follows the previously published method of Walter et al.\cite{Walter:2011gf}, where the sample is heated to $\sim$ 200 $^{\circ}$C in the presence of a XeF$_2$ crystal and a sacrificial Mo plate. The sample is then sonicated for several hours in ethanol to remove a molybdenum oxide layer formed during the process leaving a fluorine intercalated graphene layer. To intercalate the Au approximately 10 monolayers of Au atoms was deposited on the $6\sqrt{3}$ C-SiC(0001) samples which were then annealed to $\sim$ 600 $^{\circ}$C for $\sim$ 3 minutes, similar to the approach of Gierz et. al\cite{Gierz:2010cq}.

ARPES was performed on all four samples at the Electronic Structure Factory endstation (SES-R4000 analyzer) at beamline 7 of the Advanced Light Source, Lawrence Berkeley National Laboratory. A photon energy of 95 eV was used for photoexcitation with an overall energy resolution of $\sim$ 25 meV and momentum resolution of $\sim$ 0.01 \AA. The graphene on $6\sqrt{3}$ C-SiC(0001), H-SiC and F-SiC intercalated samples were transported through air to the experimental chamber and annealed to $\sim$ 500 $^{\circ}$C, $\sim$ 400 $^{\circ}$C and $\sim$ 500 $^{\circ}$C respectively to remove surface contaminants. The graphene on Au-SiC samples were prepared in a connected vacuum chamber prior to measurement and annealed to $\sim$ 300 $^{\circ}$C. During the measurements the samples were cooled to $\sim$ 20 K using a liquid He-cooled cryostat and the pressure was $<$  2 x10$^{-10}$ Torr.

Electron doping was induced in all four samples by depositing small amounts of potassium onto the surface from an SAES getter source. Charge is transferred from the randomly located potassium atoms to the carbon atoms in the graphene lattice, leading to an increase in the electron doping of the graphene. Addition of progressively more potassium atoms allows for any doping level between the starting doping level and a limit of $\sim 6 \times 10^{13} $ e/cm$^2$. Above this limit the density of potassium atoms cluster into a disordered version of the K 2 x 2 structure\cite{Bennich:1999ux} which provides little doping to the graphene and is sufficient to significantly scatter the outgoing electrons leading to a broadening of the graphene $\pi$ bands.

\section{Results and Discussion}

\subsection{Spectral function of clean samples}
The experimental spectral functions for the four samples (i - iv) considered are shown in Fig. \ref{ARPES_before} A, for the \kk\ direction. Renormalization of the bands by phonons in each case is indicated by the increase in spectral intensity, and a corresponding "kink" in the bands between \EF\ and $\sim$ 200 meV\cite{McChesney:2008p261}. For the n doped spectra (graphene on $6\sqrt{3}$ C-SiC sample, Fig. \ref{ARPES_before} A.i) it was previously assumed that the bands below and above the Dirac crossings are offset by an energy shift, \dE\ \cite{Bostwick:2007p247,Park:2009ku}. This offset causes a "kink" at the Dirac crossing ( $\sim$ -0.5 eV). A significant variation in the position of the Dirac energy below (Fig. \ref{ARPES_before} A.i) and above (Fig. \ref{ARPES_before} A.ii, A.iii and A.iv) the Fermi energy is evident for p and n doping of the graphene respectively, with hole doping increasing from left to right.

The doping variation is also clear from the size of the experimental Fermi surfaces presented in Fig. \ref{ARPES_before} B.i - B.iv. The Dirac cone picture described above (Fig. \ref{Dirac_cones} ) implies that the Fermi surface in each case should be circular, this is only the case for the Au-SiC (Fig. \ref{ARPES_before} B.ii) and H-SiC (Fig. \ref{ARPES_before} B.iii) samples. However, the Dirac picture of the bandstructure is only valid close to the $K$ point, i.e. for the small Fermi surfaces in Fig. \ref{ARPES_before} B.ii and B.iii. As the momentum values increase away from the K point the three-fold symmetry of the Brillouin zone around the K point results in the trigonally warped Fermi surfaces observed in Fig. \ref{ARPES_before} B.i and B.iv. 

The anisotropy in the Fermi surface intensity has also been studied \cite{MuchaKruczynski:2008p214,Bostwick:2007p252} and is related to the chiral nature of the electron states in graphene on account of the two sublattices. The extinguishing of intensity in the negative (positive for n-doped samples) \ky\ direction is strongly affected by symmetry breaking, as is seen in bi-layer graphene samples \cite{MuchaKruczynski:2008p214,Ohta:2006p260}. Therefore the lack of intensity in the negative (positive for n-doped samples) \ky\ direction in the current samples rules out strong symmetry breaking due to the substrate \cite{MuchaKruczynski:2008p214,Bostwick:2007p252} and the associated formation of a gap from causing the observed offset of the upper and lower bands as was proposed previously \cite{Zhou:2007p4898}. This approximation has been shown to be valid at the high photon energies employed here \cite{Shirley:1995p2916,Bostwick:2007p265,Gierz:2011do}.

\subsection{Influence of electron doping on the spectral function}
Let us recall that the separation between the hole and plasmaron bands is constant as a function of chemical doping when normalised to \ED\ and \kF\ . Here we analyze this separation for four different substrates as a function of doping. In Fig. \ref{ARPES_after} we present the experimental spectral functions from each of the samples, doped to n$_{\mathrm{e}} \sim$ 6 x 10$^{13}$ cm$^{-2}$ by the addition of potassium to the surface in the \kk\ (Fig. \ref{ARPES_after} A) and \gk\ directions( Fig. \ref{ARPES_after} B), with energy and momentum scaled to \ED\ and \kF\ respectively. Line shape analysis is performed by non-linear least squares fitting of the spectral function at fixed wave vector, $k$, for which an example is given in Fig.\ref{ARPES_after} C ($k$=0) for each of the samples. The results of peak fitting to the hole (upper) and plasmaron (lower) bands are represented by the red and black dotted lines respectively. The constant energy (binding energy ( \EBind\ )  = \ED\ ) and constant momentum ( $k$ = 0) intensity maps from spectra similar to Fig.\ref{ARPES_after} B are plotted as a function of potassium doping in Fig. \ref{EDC} i and ii. The normalized width of the spectra, corresponding to the energy and momentum width of the "diamond" region, in Fig. \ref{EDC} are remarkably constant except at low doping where the scaling relationship is expected to break down and phonon interactions are expected. As this scaling relation is unique to Coulombic electron-electron interaction effects this provides strong support for the model.

Representative Energy Distribution Curves ($k/|k_\mathrm{F}| =0$) obtained from the spectral functions in Fig \ref{ARPES_after} B are presented in Fig \ref{ARPES_after} C. Lorentzian- Gaussian peaks are fitted to the profiles indicating that two bands are observed in all cases. The Lorentzian linewidth for each peak is fixed at $\sim$ 0.1, while all other parameters are fitted. The separation of the two peaks is obvious in the fluorine and hydrogen intercalated samples, Fig. \ref{ARPES_after} C iii and iv, however it is clear that a second component is needed to arrive at a reasonable description in the case of the Au intercalated and $6\sqrt{3}$ C samples. The positions of the peaks determined from these fits are overlayed on Fig.\ref{ARPES_after} A and B in red (hole band) and black (plasmaron band) and provide a good description in all  four cases. At high binding energy only the plasmaron band is observed, the hole band emerges from the plasmaron band close to the lower Dirac crossing, and the two bands are separated. After passing through all three Dirac crossings the plasmaron band curves back up to meet the hole band, with the separation of the bands at the Dirac crossing, \Eb, being different for each of the samples. The amplitude (circles) and Gaussian linewidth (crosses) of the fitted peaks are shown in Fig. \ref{Fit_params}. We observe a decrease in the amplitude, and corresponding increase in the linewidth, of the plasmaron band with increasing momentum. The reverse is true for the hole band. This behaviour is predicted by the  \RPA\  plasmaron theory of Polini et al. and Hwang et al.  \cite{Polini:2008p3589,Hwang:wh,Hwang:to,Hwang:2008p3441} and provides further support to the existence of plasmaron bands in graphene on all four substrates.

There is strong debate about the source of the deviation of the upper and lower bands in graphene on $6\sqrt{3}$ C-SiC, with both the plasmaron bands\cite{Bostwick:2007p247} and the existence of a symmetry breaking-induced gap\cite{Qi:2010p4916,Zhou:2007p4898,Kim:2008p4920} proposed. Our work shows that the variation of the shape of the bands in Fig. 3 can be accounted for purely by electron-electron interactions (In any case, symmetry breaking would be expected to be very weak for graphene on Au, H, or F). Qi et al. \cite{Qi:2010p4916} have presented a defect model for the structure of the $6\sqrt{3}$ C layer based on defects observed in STM data, and another defect based model was presented by Kim et al \cite{Kim:2008p4920}. In the Qi model\cite{Qi:2010p4916} a significant deviation from the structure of defect free graphene is proposed and therefore a large potential barrier would need to be overcome to transform the $6\sqrt{3}$ C layer into defect free graphene. This is in contrast to the relative ease with which atomic layers are intercalated under the $6\sqrt{3}$ C layer (eg. Au \cite{Gierz:2008p3404} , F\cite{Walter:2011gf}  and H \cite{Riedl:2009p4922} presented here) resulting in essentially defect free graphene.

Qi et al  \cite{Qi:2010p4916} and Kim et al \cite{Kim:2008p4920} propose a structure for the $6\sqrt{3}$ C layer, based on STM measurements, and use this to calculate a band structure for both the $6\sqrt{3}$ C layer and the graphene on the $6\sqrt{3}$ C layer. Contributions to the electronic structure from the $6\sqrt{3}$ C layer close to the Dirac energy are used to explain the apparent gap in the ARPES spectra of graphene on $6\sqrt{3}$ C-SiC. The experimental spectral function of the buffer layer has been investigated previously\cite{Emtsev:2008p4975} and none of the detailed ARPES spectral features predicted by the models of Qi et al\cite{Qi:2010p4916} or Kim et al\cite{Kim:2008p4920} were observed. Importantly, neither model correctly predicts the insulating behaviour of the $6\sqrt{3}$ C layer, with no states above -0.4 eV binding energy observed experimentally\cite{Emtsev:2008p4975}. Defect models cannot, therefore, explain the experimental observations of either the $6\sqrt{3}$ C layer or graphene on $6\sqrt{3}$ C- SiC, and the deviation from linear behaviour at \ED\ is due to the formation of plasmaronic bands in all four of the samples presented here, including graphene on $6\sqrt{3}$ C-SiC.

\subsection{Effective screening analysis}
In Fig. \ref{alpha} the variation of the width and height of the diamond region ( \dE\ and \dk\  respectively) with  graphene effective coupling constant, \AG\ , as determined from \RPA\ theory is shown. These values were determined by generating a theoretical spectral functions for a range of \AG\ values and fitting peaks to determine the height and width of the diamond region. The experimental widths, heights and the corresponding coupling constant (derived from these by comparison with theory) are indicated by dashed lines and recorded in Table \ref{eff_screen}. As described in section \ref{sec.background} the effective coupling constant can be used to determine the dielectric constant, $\epsilon$, and an upper limit to the substrate effective dielectric constant, $\epsilon_{\mathrm{S}}$, which are also listed in Table \ref{eff_screen}. The metallic substrate, Au, provides the highest effective screening ($\epsilon_{\mathrm{S}}\approx $87), an order of magnitude larger than F ($\epsilon_{\mathrm{S}}\approx $10) or H ($\epsilon_{\mathrm{S}}\approx $7.8). Interestingly the supposedly electronically decoupled $6\sqrt{3}$ C layer ($\epsilon_{\mathrm{S}}\approx $43) provides a relatively high effective screening. This can be attributed to the presence of dangling Si bonds, which are quite polarizable \cite{Bostwick:2010p3387}.

\begin{table}[t]
\caption{Effective screening constants for Graphene determined from experimental spectral functions: Energy ( \dE\ ) and momentum ( \dk\ ) separation of the hole (upper) and plasmaron (lower) bands, graphene effective coupling constant ( $\alpha _{\mathrm{G}}= e^2/ \epsilon \hbar v_{\mathrm{F}} \sim 2.2/\epsilon$ ), the graphene effective dielectric constant ($\epsilon$) and the substrate effective dielectric constant ($ \epsilon_{\mathrm{S}} \approx 2\epsilon - 1$)}
\label{eff_screen}\centering%
\begin{tabular*}{8.5cm}{p{1.5cm}p{1.4cm}p{1.4cm}p{1.3cm}p{1.1cm}p{1.5cm}}
\hline%
Substrate			& \dE\	& \dk		& \AG\	& $\epsilon$	& $\epsilon_{\mathrm{S}}$	\\\hline%
Au-SiC			&0.12$\pm$0.04	&0.09$\pm$0.02	&0.05$\pm$0.01	&44$\pm$9	&87$\pm$18				\\
$6\sqrt{3}$ C-SiC	&0.21$\pm$0.02	&0.16$\pm$0.01	&0.1$\pm$0.04		&22$\pm$	8	&43$\pm$16				\\
F-SiC			&0.40$\pm$0.05	&0.29$\pm$0.03	&0.4$\pm$0.05		&5.5$\pm$0.7	&10$\pm$1.3				\\ 
H-SiC			&0.49$\pm$0.02	&0.34$\pm$0.01	&0.5$\pm$0.03		&4.4$\pm$0.3	&7.8$\pm$0.5				\\\hline%
\end{tabular*}
\end{table}

\subsection{Comparison to \RPA\ theory}
In order to provide a detailed comparison between theory and experiment we present in Fig. \ref{alpha_theory} the calculated spectral functions on the basis of the \RPA\ theory of Polini et al.\cite{Polini:2008p3589} in the \kk\ ( \bf A \rm) and \gk\ ( \bf B \rm) directions using the experimentally determined alpha values. Overlayed are the experimental fitted hole (red) and plasmaron (black) bands obtained for Graphene on Au-SiC, i, Graphene on $6\sqrt{3}$ C-SiC, ii, Graphene on F-SiC,iii, and Graphene on H-SiC, iv. In all cases good agreement is found between theory and experiment above the high energy Dirac crossing, \Ec\ . Below \Ec\ the experimental hole and plasmaron bands quickly merge in contrast to the prediction of the \RPA\ theory. The experimental intensity (see Fig. \ref{ARPES_after}) in this high energy region is also poorly described by the theory indicating that further theoretical work needs to be undertaken to accurately describe the spectral function below \Ec\ .

The data presented indicate that the separation of the hole and plasmaronic bands can be modified by the use of various substrates, which alter the effective dielectric coupling in the graphene. Doping of graphene is possible without exerting an influence on the band separation by the addition of potassium to the surface, thereby allowing the plasmaronic and electronic band structures to be separately manipulated.

\section{conclusion}

The shape of the spectral function of graphene on SiC(0001) and the presence of side bands to the linear bands near the K point of the Brillouin zone has recently been interpreted in terms of the formation of coupled hole-plasmon quasiparticles (ÒplasmaronsÓ). Here we study graphene, on four different interface structures (the $(6\sqrt{3}\times6\sqrt{3}) R30^{\circ}$ reconstruction, and the SiC - graphene interface intercalated with gold, hydrogen or fluorine), using angle-resolved photoemission. Our data show that similar side bands are found in all cases as revealed by energy and momentum distribution curve line shape analysis. These show the scaling of the separation with the energy of the Dirac point and the Fermi wave vector. The separation of the plasmaronic side bands and the bands unaffected by hole-plasmon coupling is found to strongly vary with interface structure. We use \RPA\ calculations to determine an upper limit to the effective dielectric constant of the underlying substrate interfacial layer; this is found to vary from $\epsilon_{\mathrm{S}} \sim$7.8 for hydrogen to $\epsilon_{\mathrm{S}} \sim$87 for Au intercalation. Graphene is thus an ideal candidate for investigating the effective screening in the context of hole-plasmon interactions. We also show that plasmaronic and electronic properties of graphene can be separately manipulated.

\begin{acknowledgments}
The Advanced Light Source is supported by the Director, Office of Science, Office of Basic Energy Sciences, of the U.S. Department of Energy under Contract No. DE-AC02-05CH11231.Work in Erlangen was supported by the ESF and the DFG through the EUROCORES program EUROGRAPHENE. A.W. acknowledges support by the Max Planck Society.
\end{acknowledgments}



%
%

%


\bibliography{references}

\end{document}